# Features of Hypothetical Plasma Phase Transition in Interiors of Saturn and Jupiter

*Ukrainets A.V., Iosilevskiy I.L.*

*Moscow Institute of Physics and Technology (State University), Moscow*
*ukrainets_art@mail.ru, ilios@orc.ru*

Here we consider physics of phase transitions in the matter under extreme conditions, which corresponds to giant planets, brown dwarfs and some other substellar astrophysical objects. We examine the anomalous features of one of the hypothetical Plasma Phase Transitions (PPT) [1], namely the PPT of Saumon & Chabrier [2], which is expected to occur in the hydrogen-helium plasma in the interior of Jupiter and Saturn [3]. Many of available theories [3] use the so-called "additivity rule" in constructing the equation of state of for hydrogen-helium mixture. In frames of this approximation specific volume and specific enthalpy of the mixture at given pressure and temperature (*P,T*) are defined as a simple sum of specific volumes and enthalpies of the components at the same *P* and *T*. It has been stated previously [4] that in frames of "additivity rule" all phase transition boundaries in *P-T* coordinates for both mixed components are directly transferred without any change onto the final *P-T* phase diagram of the mixture. Thus, in *P–T* representation the structure of the final PPT boundaries for helium-hydrogen mixture in additivity approximation is equivalent to the simple superposition of PPT boundaries from purely hydrogen [Ebeling et al] and purely helium [Ebeling et al] [Winisdoerffer & Chabrier] plasma to the hydrogen-helium mixture [2]. At the same time having investigated the *non-congruent* nature of the phase equilibrium in the high-temperature mixtures of two or more chemical elements [4], one expects to find a *fundamentally different* character of the phase boundaries in the hydrogen-helium plasma of the astrophysical objects.

In this work using tabular data [2] we reconstructed some thermodynamic quantities of the hydrogen-helium plasma in the range of parameters corresponding to the conditions at the co-existence boundaries of the PPT version [2]. Using general thermodynamic relations we also reconstructed the characteristics of the Coulomb and density corrections (the so-called non-ideality corrections). That allowed us to calculate two previously unknown characteristics of the studied PPT: (**A**) the jump of the electrostatic potential, which is generally inherent in any phase transition in equilibrium Coulomb systems [5], at the phase boundary of the PPT [2], and (**B**) typical scale of the PPT *non-congruency* value (the differences in chemical composition of the coexisting phases). While the first effect – the potential of phase boundary – turned out to be within the typical scale of contact electrochemical quantities [5], $\Delta\varphi \sim 1\text{-}2$ eV (Fig.1), the second quantity – the PPT non-congruency – is rather considerable under conditions inside Saturn and Jupiter.

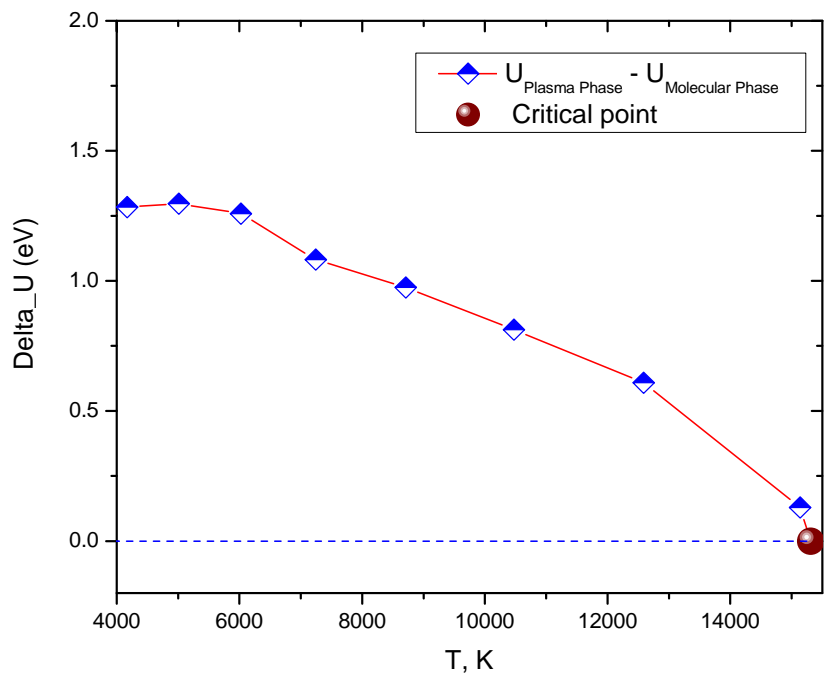

**Fig. 1**. Presently estimated electrostatic potential of phase boundary for hypothetical plasma phase transition (PPT) of Saumon & Chabrier [2].

The value and the sign of the non-congruency (the latter is very important) are in good agreement with the experimentally observed helium deficiency in the atmosphere of giant planets [6] (see Fig. 2). According to the review [6] Y(He)_Jupiter ≈ 0.231, Y(He)_Saturn ≈ 0.215. It should be noted that one of the aims of the Cassini-Huygens apparatus, which is now on the orbit of Saturn, is to check and to derive more accurately the above mentioned helium deficiency in the Saturn atmosphere.

The next planned step of present study is to reconstruct the value of the splitting of the boiling curve $P_{boil}(T)$ and the saturation curve $P_{satur}(T)$ on one $P$-$T$ diagram, which is typical of the non-congruent phase transitions [4]. Another planned direction to continue present study is to provide the same reconstruction procedure of the electrostatics and the non-congruency for many other numerous versions of PPT in hydrogen-helium mixture described in the literature (for example, [7]), including the PPT, which originates in the computations conducted using the "chemical plasma model" implemented in SAHA-IV [8, 9] at a definite combination of the interaction parameters of the particles composing the hydrogen-helium mixture.

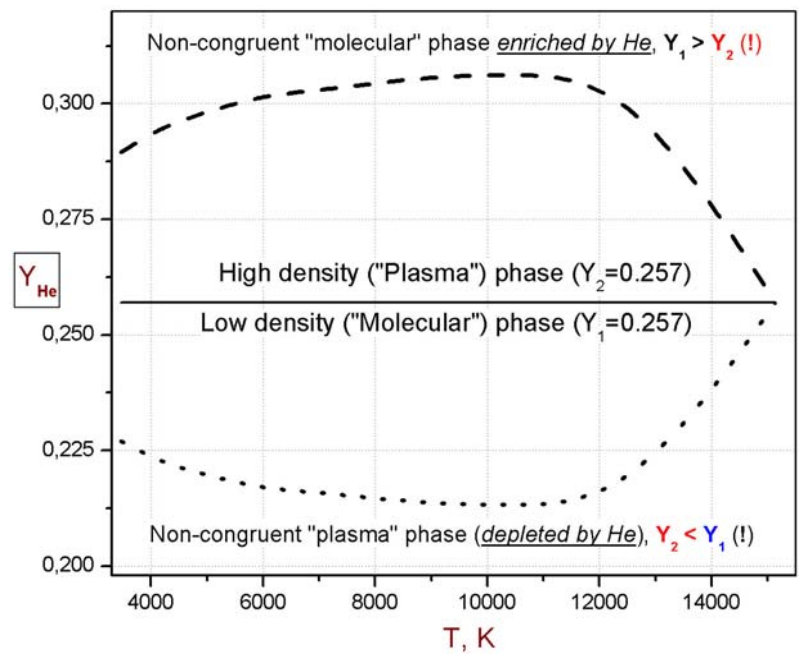

**Fig. 2.** Non-congruence of hypothetical plasma phase transition (Saumon & Chabrier version) in hydrogen-helium mixture in interiors of Jupiter and Saturn ($Y_{He}$ ~ 0.257). Estimated helium abundance vs. temperature for the coexisting "plasma" and "molecular" phases.

Present work is supported by Grant CRDF (REC MO-011-0), by RAS Scientific Research Programs: "Thermophysics of the intensive energy impacts" and "Physics of the matter under high pressure in interior of planets".